\documentclass[aps,twocolumn,amsmath,amssymb,reprint,numbers,superscriptaddress,noeprint,longbibliography]{revtex4-1}

\usepackage{braket}
\usepackage{soul}
\usepackage{geometry}
\usepackage{graphicx}
\usepackage{booktabs}
\usepackage{array}
\usepackage{relsize}
\usepackage{color}
\usepackage{xcolor}
\usepackage{eqnarray}
\usepackage{caption}
\usepackage{textcase}
\usepackage{textcomp}
\usepackage{setspace}
\usepackage{xcolor}
\usepackage{placeins}
\usepackage{subfig}
\usepackage{tikz}
\usetikzlibrary{positioning}
\usepackage{hyperref}
\hypersetup{
     colorlinks   = true,
     linkcolor    = blue,
     citecolor    = blue,
     urlcolor = blue
}

\definecolor{cola}{rgb}{0.7,0.1,0.1}
\definecolor{colb}{rgb}{0.9,0.4,0}
\definecolor{colc}{rgb}{0.3,0.7,0}
\definecolor{cold}{rgb}{0,0.35,0.75}
\definecolor{cole}{rgb}{0.63, 0.13, 0.94}
\definecolor{colf}{rgb}{0.5, 0.5, 0.5}

\begin{document}

\title[]{Coherent Spin Preparation of Indium Donor Qubits in Single ZnO Nanowires}

\author{Maria L. K. Viitaniemi}
\email{mariav93@uw.edu}
\affiliation{Department of Physics, University of Washington, Seattle, Washington 98195, USA}
\author{Christian Zimmermann}
\affiliation{Department of Physics, University of Washington, Seattle, Washington 98195, USA}
\author{Vasileios Niaouris}
\affiliation{Department of Physics, University of Washington, Seattle, Washington 98195, USA}
\author{Samuel H. D'Ambrosia}
\affiliation{Department of Physics, University of Washington, Seattle, Washington 98195, USA}
\author{Xingyi Wang}
\affiliation{Department of Electrical Engineering, University of Washington, Seattle, Washington 98195, USA}
\author{E. Senthil Kumar}
\affiliation{Department of Physics, Simon Fraser University, Burnaby, British Columbia V5A 1S6, Canada}
\author{Faezeh Mohammadbeigi}
\affiliation{Department of Physics, Simon Fraser University, Burnaby, British Columbia V5A 1S6, Canada}
\author{Simon P. Watkins}
\affiliation{Department of Physics, Simon Fraser University, Burnaby, British Columbia V5A 1S6, Canada}
\author{Kai-Mei C. Fu}
\affiliation{Department of Physics, University of Washington, Seattle, Washington 98195, USA}
\affiliation{Department of Electrical Engineering, University of Washington, Seattle, Washington 98195, USA}

\begin{abstract}
Shallow donors in ZnO are promising candidates for photon-mediated quantum technologies. Utilizing the indium donor, we show that favorable donor-bound exciton optical and electron spin properties are retained in isolated ZnO nanowires. The inhomogeneous optical linewidth of single nanowires (60\,GHz) is within a factor of 2 of bulk single-crystalline ZnO. Spin initialization via optical pumping is demonstrated and coherent population trapping is observed. The two-photon absorption width approaches the theoretical limit expected due to the hyperfine interaction between the indium nuclear spin and the donor-bound electron.
\end{abstract}

\date{\today}

\maketitle


Shallow donor spin qubits, composed of an electron bound to a donor ion, are one of the simplest solid-state qubit systems and have the potential for ultra-long qubit coherence times. For example, donors in silicon have demonstrated qubit coherence times from seconds~\cite{tyryshkin2012esc} to minutes~\cite{saeedi2013rtq}.
Unlike silicon, shallow neutral donors (D$^0$) in direct bandgap semiconductors exhibit efficient optical coupling to donor-bound excitons (D$^0$X). This coupling allows for the transfer of quantum information between the electron spin state and a photon, hence enabling photon-based applications in quantum communication~\cite{Wehner2018qiv} and computation~\cite{Benjamin2009pmb}.
The direct bandgap semiconductor ZnO is a particularly attractive host due to its large exciton binding energy~\cite{Meyer2004bed}, low spin-orbit coupling~\cite{Khaetskii2001sft} and potential for a nuclear spin-free host with Zn isotope purification. Al, Ga, and In substituting for Zn are common shallow donors in ZnO, with In having the largest binding energy of the three~\cite{Meyer2004bed}.

Most quantum applications utilizing such optically-active donors will require the isolation of single donors~\cite{Ladd2010qc,Benjamin2009pmb} and nanoscale device integration~\cite{Schmidgall2018fcs,Schroder2016qnd}. However, obtaining good spin and optical properties for donors in nanostructures may be challenging; the extended effective-mass wave function of the donor and the resulting shallow donor binding energy leave the donor sensitive to surface noise~\cite{Kane1998sbn,Ramdas1981sss}. Here, we show the promise of utilizing a bottom-up technique to isolate a small ensemble of In donors. 
We demonstrate that the inhomogeneous In D$^0$X linewidth in ensembles of nanowires (20\,GHz) is comparable to the bulk single-crystalline D$^0$X linewidth (15-25\,GHz). Dropcasting to isolate single nanowires only increases the linewidth to 60\,GHz. These narrow optical linewidths enable the optical probing of the spin properties of the donor ground-state. 
In single nanowires, we demonstrate spin initialization into both ground spin states via optical pumping~\cite{Linpeng2016lsr, Sleiter2013ops} and preparation of a coherent superposition spin state via coherent population trapping~\cite{Fu2005cpt, Santori2006cpt, Gray1978cta, Xu2008cpt}. The measured 1\,GHz two-photon absorption width approaches the limit expected due to the hyperfine interaction of the ground-state electron with the In spin-9/2 nuclear spin~\cite{Gonzalez1982mrs, Block1982odm}.


ZnO nanowire samples are grown on a \textit{c}-plane sapphire (0001) substrate via metal organic chemical vapor deposition (MOCVD) as described in \cite{kumar2013oed}. 
Nanowires are typically 100-200\,nm in diameter,  1-4\,\textmu m long, and grow in a dense ensemble. The ZnO crystal \textit{c}-axis [0001] points along the long axis of the nanowire. A scanning electron micrograph (SEM) image of the nanowire ensemble is shown in inset~(i) of Fig.~\ref{Fig1}c and in the supplementary material~\cite{SI}.
To isolate single nanowires, the ensemble sample is sonicated in ethanol to detach the nanowires from the sapphire substrate. The nanowires are then dropcast from the solution onto a SiO$_2$ substrate. 
An SEM image of a single dropcast nanowire is shown in inset~(ii) of Fig.~\ref{Fig1}c. 
Isolation of single nanowires is achieved via scanning optical microscopy with representative laser reflection and photoluminescence images shown in Fig.~\ref{Fig1}b. 

The nanowires are n-type due to unintentional doping during growth. In and Ga D$^0$X lines are both present in the photoluminescence spectra (Fig.~\ref{Fig1}c); here, we focus on In due to its larger donor binding energy and higher concentration.
The In donor concentration is estimated to have an upper limit of 10$^{16}$~cm$^{-3}$ from nanoprobe resistivity measurements on similar samples.

Samples are mounted in a helium immersion cryostat with a superconducting magnet. Measurements are performed at 5.2\,K. 
The magnetic field is aligned to the optical axis $\hat{k}$ ($\hat{k} \parallel\vec{B}$). 
Ensemble measurements are performed with the optical axis  parallel to the crystal axis ($\hat{k} \parallel\hat{c}$). 
Because the single dropcast nanowires lie horizontally on the substrate, single nanowire measurements are performed with the optical axis (and magnetic field axis) perpendicular to the crystal axis ($\hat{c} \perp \hat{k},\vec{B}$).
We utilize a confocal microscope with a lateral point spread function of $\sim$1\,\textmu m. In the ensemble sample, this corresponds to approximately five nanowires on a floor of partially nucleated nanostructures (Fig.~\ref{Fig1}c inset~(i)). In the single nanowire samples, we estimate that ensembles of a couple hundred indium donors are simultaneously addressed.
\begin{figure*}[t]
  \centering
  \includegraphics[width=7in]{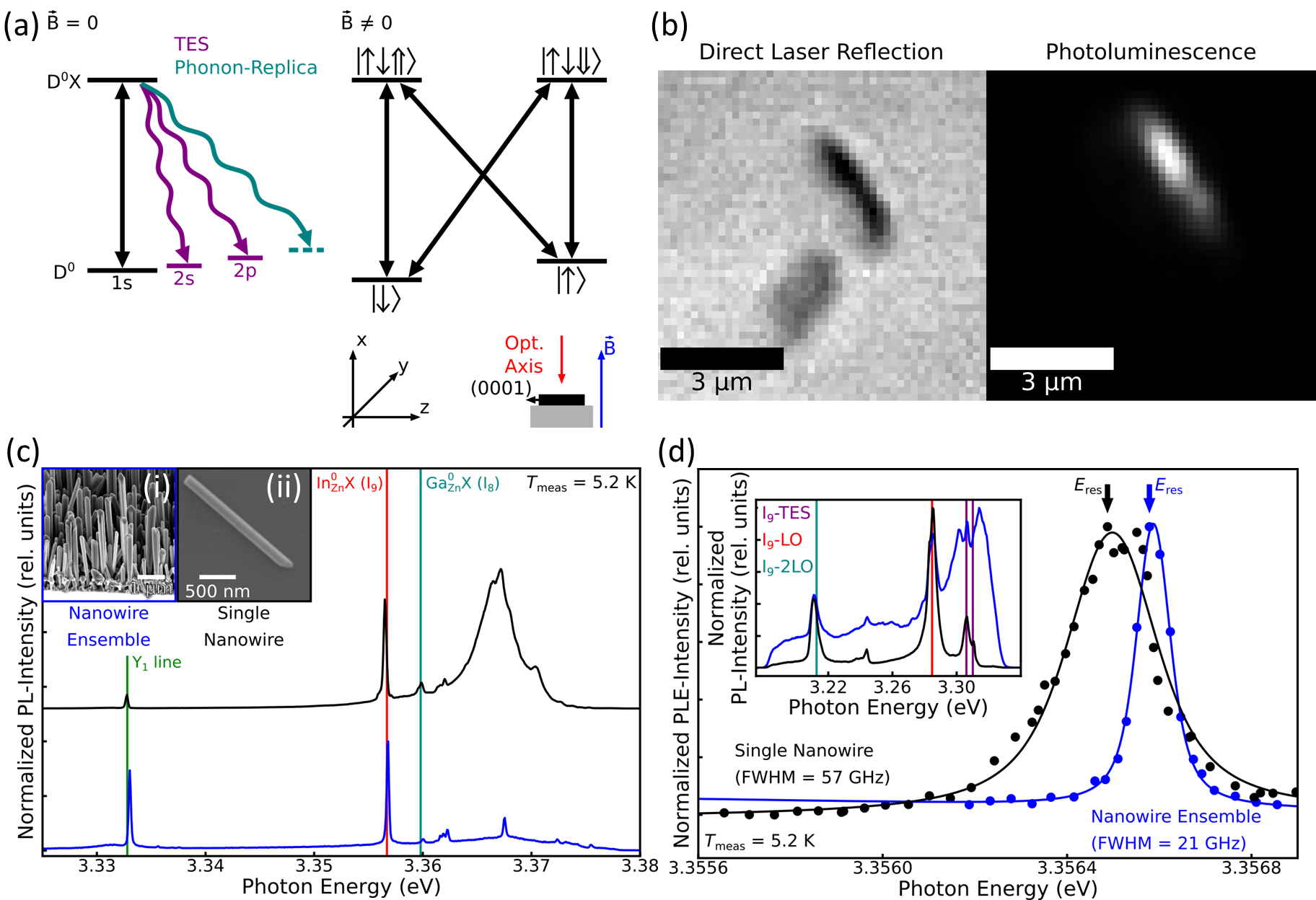}
  \caption{\label{Fig1}  
  (a)~Energy level diagram of the D$^0$ and D$^0$X system. The $\ket{\uparrow}$ ($\ket{\Uparrow}$) denotes the electron (hole) spin state. The diagram shows the experimental geometry for a single nanowire.
  (b)~Confocal images of a single nanowire. Left panel: laser reflection is collected. Right panel: In D$^0$X TES and LO-phonon replicas are collected under resonant excitation of the In D$^0$X transition.
  (c)~Spectra of an ensemble (blue) and single (black) nanowire with above-band gap excitation at 3.4440\,eV, 5.2\,K, and 0\,T. The excitation power is 10\,nW (36\,nW) for the ensemble (single) nanowires. The single nanowire spectra has been offset for clarity. Labeled transitions are from~\cite{Meyer2004bed}. The inset shows SEM images of an (i) ensemble and a (ii) single nanowire.
  (d)~In D$^0$X PLE spectra of an ensemble (blue) and single (black) nanowire at 5.2\,K and 0\,T. 
  The solid lines are Voigt fits to the data. The inset shows the TES and LO-phonon replica spectrum with the excitation laser resonant on the In D$^0$X transition.
  For the nanowire ensemble PLE, the broad non-specific background is subtracted from the spectrum before summing.
  }
\end{figure*}

The energy level diagram of the In donor system is shown in Fig.~\ref{Fig1}a. 
The magnetic field lifts the spin degeneracy of the D$^0$ and the D$^0$X states due to the electron and hole Zeeman effects~\cite{Ding2009cbe, wagner2009gvb, SI}, respectively. This leads to four D$^0$-D$^0$X transitions with nominally two polarized in each $\hat{\textrm{y}}$ and $\hat{\textrm{z}}$; however, these polarization rules are relaxed when coupling into the end of a nanowire.
Moreover, the system may relax via several lower energy transitions such as the longitudinal-optical (LO) phonon replicas and the two electron satellite (TES) transitions (corresponding to relaxation to an excited hydrogenic D$^0$ orbital)~\cite{Meyer2004bed}.

\begin{figure}[h]
    \centering
    \includegraphics[width = 3.5in]{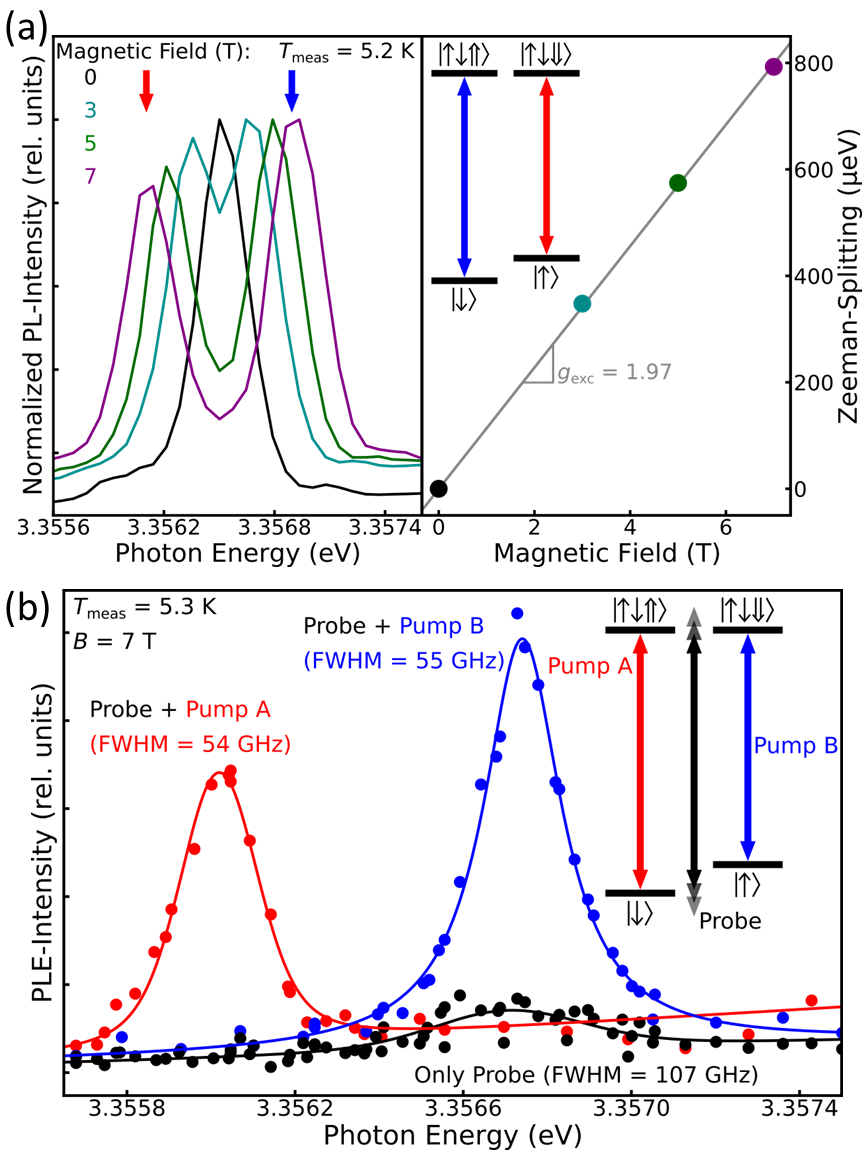}
    \caption{
    (a)~Magnetic field-dependent photoluminescence spectra of the In D$^0$X transitions for a single nanowire at 5.2\,K. Data are shown at 0 (black), 3 (teal), 5 (green), and 7\,T (purple). The right panel shows the Zeeman splitting as a function of magnetic field. The gray line is a fit to the data which yields the sum of the electron and hole $g$-factors.
    (b)~Single nanowire PLE spectra of the In D$^0$X transitions at 7\,T, 5.3\,K with no pump (black), pump on the high energy transition (red), and pump on the low energy transition (blue) are shown. All laser powers are 7.2\,\textmu W. The two-laser datasets (red and blue) are offset for ease of comparison. 
    The CPT dip is not visible because of the low spectral resolution.}
    \label{Fig2}
\end{figure}



Photoluminescence spectra of both the ensemble and single nanowire samples are shown in Fig.~\ref{Fig1}c. 
The In D$^0$X transition is observed at 3.3567\,eV. In addition to the In, Ga and Y-line donor-bound exciton features~\cite{Meyer2004bed}, the broad surface exciton is observed between 3.365-3.370\,eV~\cite{Wischmeier2006dse, Biswas2011mos, Grabowska2005see, Travnikov1989sez}.
The inset in Fig.~\ref{Fig1}d shows the In D$^0$X TES and LO-phonon replicas under resonant In D$^0$X excitation. 
A striking contrast can be observed between the two samples. In the ensemble sample, the TES and LO-phonon replicas are on a large background which is uncorrelated with the In D$^0$X transition. This background may be related to the thin 2-dimensional layer of ZnO on the ensemble substrate as well as to the observed non-uniformity of individual nanowires.
For single nanowire measurements, nanowires are screened for sharp In D$^0$X transitions and low non-specific background in the TES/LO-phonon replica region~\cite{SI}.

The In D$^0$X linewidth can be spectrometer-resolution limited. To obtain higher-resolution spectra, we perform photoluminescence excitation spectroscopy (PLE). In PLE measurements, D$^0$X TES and LO-phonon replicas are monitored while tuning the excitation laser wavelength over the In D$^0$X transitions. For the nanowire ensembles, the non-specific background is subtracted and only the LO-phonon replicas are monitored.
The excitation power is 200\,nW (7.2\,\textmu W) for the ensemble (single) nanowires. We are able to use a lower excitation power for the ensemble nanowires, because of the larger volume of material within our measurement spot and the increased collection efficiency from the top of the nanowires due to wave-guiding (see supplementary material~\cite{SI}).
As shown in Fig.~\ref{Fig1}d, in nanowire ensembles, PLE linewidths of 20\,GHz are measured. These are comparable to the narrowest linewidths (15-25\,GHz) measured in our lab on bulk single-crystalline ZnO for Ga D$^0$X transitions at 5\,K~\cite{SI}. In single nanowires, linewidths as narrow as 57\,GHz are observed.
We attribute the increase in linewidth to additional strain caused by the dropcasting process as well as spectral diffusion and several homogeneous broadening mechanisms we will discuss further below.
Due to the high non-specific background in the ensemble TES/LO-phonon replica region as well as the ability to isolate fewer donors in the isolated nanowires, we focus on single nanowire measurements for the remainder of this work.

In order to utilize the D$^0$ system for quantum applications, the spin degeneracy of the D$^0$ and D$^0$X states must be lifted. A magnetic field may be used to split the ground state as determined by the electron $g$-factor ($g_e$). The splitting of the excited state is then solely determined by the hole $g$-factor ($g_h$), because the excited state electrons form a spin singlet. As shown in Fig.~\ref{Fig2}a, when a magnetic field is applied perpendicular to the crystal $c$-axis, the sum of $g_e$ and $g_h^{\perp}$ is measured to be $g_{tot}=1.97$ in a single nanowire. This is consistent with reported values for $g_e$ (1.9-2.0) and the lowest values reported for $g_h^{\perp}$ (0.1-0.3)~\cite{Ding2009cbe, wagner2009gvb}. As described later in the text, we use high-resolution PLE to further confirm a $g_e$ of 1.90 and therefore a lower bound for $g_h^{\perp}$ of 0.07.

When a magnetic field is applied to the donor system, the spin degeneracy is lifted and donor spin initialization can be performed by resonant optical pumping~\cite{Linpeng2016lsr,Sleiter2013ops}. 
For example, to optically pump the system into the $\ket{\downarrow}$ state, a resonant laser may be applied to the $\ket{\uparrow} \Leftrightarrow \ket{\Downarrow \uparrow \downarrow}$ transition. Because the D$^0$X state can relax into either D$^0$ spin state~\cite{wagner2009gvb}, after several cycles, the spin is initialized to the $\ket{\downarrow}$ state.
Fig.~\ref{Fig2}b shows three PLE spectra which confirm that optical pumping can be achieved in a single nanowire. 
First, we perform a single-laser scan over the D$^0$X transitions at 7\,T (black data). Due to the small observed hole splitting, we expect the two transitions sharing the same electron spin ground-state level will not be resolved.
For these single-laser, steady-state measurements, we also expect the PLE intensity to be significantly weaker at 7\,T than it was at 0\,T due to optical pumping. For the pair of transitions involving the $\ket{\uparrow}$ state, the signal lies below the noise floor. 
To confirm that the overall low signal is due to optical pumping, we perform two-laser spectroscopy. A resonant pump laser is applied on the $\ket{\downarrow} \Leftrightarrow \ket{\Uparrow \uparrow \downarrow}$ transition while a second probe laser is scanned over all the In D$^0$X transitions (red data). In this two-laser PLE, the pump laser is able to re-pump population from $\ket{\downarrow}$ to $\ket{\uparrow}$ so that the $\ket{\uparrow} \Leftrightarrow \ket{\Downarrow \uparrow \downarrow}$ transition signal is recovered. 
A similar phenomenon can be observed by placing the pump laser on the $\ket{\uparrow} \Leftrightarrow \ket{\Downarrow \uparrow \downarrow}$ transition and recovering the $\ket{\downarrow} \Leftrightarrow \ket{\Uparrow \uparrow \downarrow}$ transition signal (blue data).
The 7\,T two-laser linewidth of 55\,GHz corresponds to the spectral diffusion and homogeneously broadened linewidth of the donor subpopulation that is optically pumped. This is within the uncertainty of the 57\,GHz 0\,T single nanowire PLE linewidth (Fig.~\ref{Fig1}d) which includes static inhomogeneous broadening from the entire donor ensemble. The similarity between the 0\,T single-laser and the 7\,T two-laser linewidths suggest broadening mechanisms, such as  
power broadening~\cite{Reimer2016opb}, laser-induced temperature broadening, and
spectral diffusion that is fast compared to our measurement time (20\,s)~\cite{Holmes2015sdi, Empedocles1999isd, MacQuarrie2021gtc}, are the dominant line broadening mechanisms in all of the single nanowire spectra. All of these mechanisms may be exacerbated by the higher powers used for single nanowire measurements relative to the ensemble nanowire measurements.

\begin{figure}[h]
    \centering
    \includegraphics[width = 3.2in]{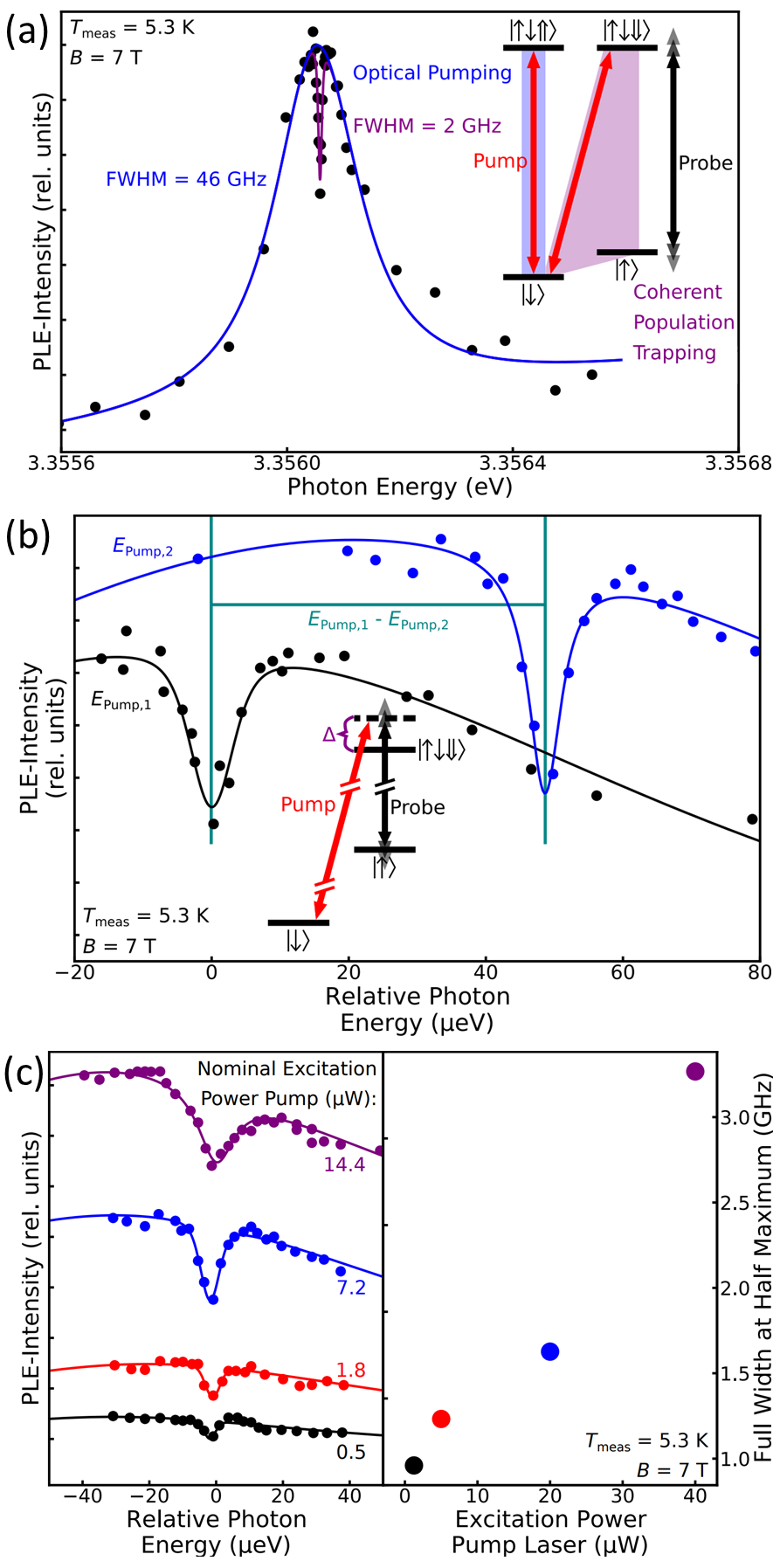}
    \caption{
    Single nanowire, high resolution PLE spectra of the In D$^0$X transitions at 7\,T, 5.3\,K. All laser powers are 7.2\,\textmu W unless otherwise noted.
    (a)~PLE showing the CPT dip.  The pump laser is on resonance with the $\ket{\downarrow} \Leftrightarrow  \ket{\Uparrow\uparrow\downarrow}$ transition.
   (b)~PLE over the CPT dip with the pump laser wavelength at 3.35687\,eV (3.35682\,eV) for the black E$_{\textrm{Pump,1}}$ (blue E$_{\textrm{Pump,2}}$) data.
   (c)~The left panel shows PLE spectra over the CPT dip with the pump laser wavelength at 3.35774\,eV. The power of the pump laser is 14.4, 7.2, 1.8, and 0.5\,\textmu W (from top to bottom).
   }
    \label{Fig3}
\end{figure}

High-resolution two-laser scans, shown in Fig.~\ref{Fig3}a, reveal a sharp 2\,GHz wide dip on the PLE peak that is the result of coherent population trapping (CPT).
In CPT, the system enters a dark state when the two transitions of a $\mathrm{\Lambda}$ system are simultaneously pumped~\cite{Fu2005cpt, Santori2006cpt, Gray1978cta, Xu2008cpt}.
For example, the $\mathrm{\Lambda}$ system can be formed between the two electron ground states and a single excited D$^0$X state.
In our experiments, because the Zeeman splitting between the D$^0$X states is small relative to the two-laser linewidth (Fig.~\ref{Fig2}b), the pump laser is able to excite both $\ket{\downarrow} \Leftrightarrow \ket{\Uparrow \uparrow \downarrow}$ and $\ket{\downarrow} \Leftrightarrow \ket{\Downarrow \uparrow \downarrow}$ simultaneously; therefore, as shown in the inset to Fig.~\ref{Fig3}a, a $\mathrm{\Lambda}$ system is formed by the pump laser driving $\ket{\downarrow} \Leftrightarrow \ket{\Downarrow \uparrow \downarrow}$ and the probe laser on $\ket{\uparrow} \Leftrightarrow \ket{\Downarrow \uparrow \downarrow}$.

To confirm that the dip is related to CPT, we detune the pump laser energy by $\Delta$ from $\ket{\downarrow} \Leftrightarrow \ket{\Downarrow \uparrow \downarrow}$ and measure the shift of the dip (Fig.~\ref{Fig3}b).
We observe a one-to-one correlation between $\Delta$ and the spectral position of the dip~\cite{SI}.
As expected, the separation between the pump energies and the dip energies (0.76-0.77\,meV) is consistent with a $g_e$ of 1.90~\cite{wagner2009gvb}, confirming a lower bound for $g_h^{\perp}$ of 0.07.
Moreover, Fig.~\ref{Fig3}c shows that reducing the pump laser power reduces the dip linewidth and contrast as expected in CPT~\cite{Fu2005cpt, Linpeng2021osc}.
The smallest linewidth (1\,GHz) approaches the linewidth expected due to the spin-9/2 In nuclear-electron hyperfine interaction, where the In line splits into 10 hyperfine lines separated by 100\,MHz each~\cite{Gonzalez1982mrs, Block1982odm}.
Moreover, considering spectral diffusion as a broadening mechanism for the two-laser peak (Fig.~\ref{Fig2}b), the narrowness of the CPT dip (1\,GHz) compared to the two-laser peak (55\,GHz) rules out ground-state spectral diffusion; therefore, excited-state spectral diffusion is the dominant spectral diffusion mechanism.

In summary, In donors in single ZnO nanowires preserve several important properties for utilizing donor spins in direct band gap materials as qubits. 
Photoluminescence excitation linewidths comparable to bulk show that even in the proximity of surfaces, high optical homogeneity is retained.
Optical spin initialization and coherent population trapping are essential steps towards the spin control needed to prepare arbitrary qubit states. We expect higher collection efficiencies and cleaner selection rules can be achieved by moving towards single nanowires in a vertical geometry~\cite{SI}. This would allow us to use lower excitation powers, minimizing the observed linewidth broadening and surface exciton. 
Additionally, there are several avenues to continue toward the single donor limit, including
increasing the purity of the ZnO precursor materials or the study of two lattice-site donor defects, like the tin complex~\cite{Kumar2016hrp}, that have lower probabilities of formation.

\section{Supporting Information}
SEMs of ensemble nanowires, 
photoluminescence spectra of additional nanowires,
description of background correction, 
0\,T photoluminescence spectra of the TES/LO-phonon replica region with resonant excitation, 
comparison of 0\,T PLE on various samples, 
magnetic field dependent spectra to measure the g-factor in ensemble nanowires, 
photoluminescence spectra of the TES/LO-phonon replica region at 7\,T with 2-laser resonant excitation, 
optical pumping demonstration of the nanowire ensemble and an additional nanowire,
low-resolution 2-laser PLE measurements with varying pump wavelength,
CPT measurements on an additional nanowire,
CPT measurements on the main text nanowire with additional pump wavelengths, and
simulation of light-guiding in a nanowire.

\section{Acknowledgements}
We would like to thank Yusuke Kozuka for the bulk ZnO samples used for comparison, Ethan Hansen for assistance in performing PLE and PL measurements on ZnO nanowire ensembles, and Xiayu Linpeng for fruitful discussions. 
The support of the Natural Sciences and Engineering Research Council of Canada is gratefully acknowledged.  
This material is based upon work primarily supported by the Army Research Office MURI Grant on Ab Initio Solid-State Quantum Materials: Design, Production and Characterization at the Atomic Scale (18057522). Bulk measurements were supported by National Science Foundation under Grant No. 1150647.
\bibliography{nanowire_paper.bib}

\begin{thebibliography}{33}%
\makeatletter
\providecommand \@ifxundefined [1]{%
 \@ifx{#1\undefined}
}%
\providecommand \@ifnum [1]{%
 \ifnum #1\expandafter \@firstoftwo
 \else \expandafter \@secondoftwo
 \fi
}%
\providecommand \@ifx [1]{%
 \ifx #1\expandafter \@firstoftwo
 \else \expandafter \@secondoftwo
 \fi
}%
\providecommand \natexlab [1]{#1}%
\providecommand \enquote  [1]{``#1''}%
\providecommand \bibnamefont  [1]{#1}%
\providecommand \bibfnamefont [1]{#1}%
\providecommand \citenamefont [1]{#1}%
\providecommand \href@noop [0]{\@secondoftwo}%
\providecommand \href [0]{\begingroup \@sanitize@url \@href}%
\providecommand \@href[1]{\@@startlink{#1}\@@href}%
\providecommand \@@href[1]{\endgroup#1\@@endlink}%
\providecommand \@sanitize@url [0]{\catcode `\\12\catcode `\$12\catcode
  `\&12\catcode `\#12\catcode `\^12\catcode `\_12\catcode `\%12\relax}%
\providecommand \@@startlink[1]{}%
\providecommand \@@endlink[0]{}%
\providecommand \url  [0]{\begingroup\@sanitize@url \@url }%
\providecommand \@url [1]{\endgroup\@href {#1}{\urlprefix }}%
\providecommand \urlprefix  [0]{URL }%
\providecommand \Eprint [0]{\href }%
\providecommand \doibase [0]{http://dx.doi.org/}%
\providecommand \selectlanguage [0]{\@gobble}%
\providecommand \bibinfo  [0]{\@secondoftwo}%
\providecommand \bibfield  [0]{\@secondoftwo}%
\providecommand \translation [1]{[#1]}%
\providecommand \BibitemOpen [0]{}%
\providecommand \bibitemStop [0]{}%
\providecommand \bibitemNoStop [0]{.\EOS\space}%
\providecommand \EOS [0]{\spacefactor3000\relax}%
\providecommand \BibitemShut  [1]{\csname bibitem#1\endcsname}%
\let\auto@bib@innerbib\@empty
\bibitem [{\citenamefont {Tyryshkin}\ \emph {et~al.}(2012)\citenamefont
  {Tyryshkin}, \citenamefont {Tojo}, \citenamefont {Morton}, \citenamefont
  {Riemann}, \citenamefont {Abrosimov}, \citenamefont {Becker}, \citenamefont
  {Pohl}, \citenamefont {Schenkel}, \citenamefont {Thewalt}, \citenamefont
  {Itoh},\ and\ \citenamefont {Lyon}}]{tyryshkin2012esc}%
  \BibitemOpen
  \bibfield  {author} {\bibinfo {author} {\bibfnamefont {Alexei~M.}\
  \bibnamefont {Tyryshkin}}, \bibinfo {author} {\bibfnamefont {Shinichi}\
  \bibnamefont {Tojo}}, \bibinfo {author} {\bibfnamefont {John~J.L.}\
  \bibnamefont {Morton}}, \bibinfo {author} {\bibfnamefont {Helge}\
  \bibnamefont {Riemann}}, \bibinfo {author} {\bibfnamefont {Nikolai~V.}\
  \bibnamefont {Abrosimov}}, \bibinfo {author} {\bibfnamefont {Peter}\
  \bibnamefont {Becker}}, \bibinfo {author} {\bibfnamefont {Hans~Joachim}\
  \bibnamefont {Pohl}}, \bibinfo {author} {\bibfnamefont {Thomas}\ \bibnamefont
  {Schenkel}}, \bibinfo {author} {\bibfnamefont {Michael~L.W.}\ \bibnamefont
  {Thewalt}}, \bibinfo {author} {\bibfnamefont {Kohei~M.}\ \bibnamefont
  {Itoh}}, \ and\ \bibinfo {author} {\bibfnamefont {S.~A.}\ \bibnamefont
  {Lyon}},\ }\bibfield  {title} {\enquote {\bibinfo {title} {Electron spin
  coherence exceeding seconds in high-purity silicon},}\ }\href {\doibase
  10.1038/nmat3182} {\bibfield  {journal} {\bibinfo  {journal} {Nature
  Materials}\ }\textbf {\bibinfo {volume} {11}},\ \bibinfo {pages} {143--147}
  (\bibinfo {year} {2012})}\BibitemShut {NoStop}%
\bibitem [{\citenamefont {Saeedi}\ \emph {et~al.}(2013)\citenamefont {Saeedi},
  \citenamefont {Simmons}, \citenamefont {Salvail}, \citenamefont {Dluhy},
  \citenamefont {Riemann}, \citenamefont {Abrosimov}, \citenamefont {Becker},
  \citenamefont {Pohl}, \citenamefont {Morton},\ and\ \citenamefont
  {Thewalt}}]{saeedi2013rtq}%
  \BibitemOpen
  \bibfield  {author} {\bibinfo {author} {\bibfnamefont {Kamyar}\ \bibnamefont
  {Saeedi}}, \bibinfo {author} {\bibfnamefont {Stephanie}\ \bibnamefont
  {Simmons}}, \bibinfo {author} {\bibfnamefont {Jeff~Z}\ \bibnamefont
  {Salvail}}, \bibinfo {author} {\bibfnamefont {Phillip}\ \bibnamefont
  {Dluhy}}, \bibinfo {author} {\bibfnamefont {Helge}\ \bibnamefont {Riemann}},
  \bibinfo {author} {\bibfnamefont {Nikolai~V}\ \bibnamefont {Abrosimov}},
  \bibinfo {author} {\bibfnamefont {Peter}\ \bibnamefont {Becker}}, \bibinfo
  {author} {\bibfnamefont {Hans-Joachim}\ \bibnamefont {Pohl}}, \bibinfo
  {author} {\bibfnamefont {John J~L}\ \bibnamefont {Morton}}, \ and\ \bibinfo
  {author} {\bibfnamefont {Mike L~W}\ \bibnamefont {Thewalt}},\ }\bibfield
  {title} {\enquote {\bibinfo {title} {Room-temperature quantum bit storage
  exceeding 39 minutes using ionized donors in silicon-28},}\ }\href@noop {}
  {\bibfield  {journal} {\bibinfo  {journal} {Science}\ }\textbf {\bibinfo
  {volume} {342}},\ \bibinfo {pages} {830--833} (\bibinfo {year}
  {2013})}\BibitemShut {NoStop}%
\bibitem [{\citenamefont {Wehner}\ \emph {et~al.}(2018)\citenamefont {Wehner},
  \citenamefont {Elkouss},\ and\ \citenamefont {Hanson}}]{Wehner2018qiv}%
  \BibitemOpen
  \bibfield  {author} {\bibinfo {author} {\bibfnamefont {Stephanie}\
  \bibnamefont {Wehner}}, \bibinfo {author} {\bibfnamefont {David}\
  \bibnamefont {Elkouss}}, \ and\ \bibinfo {author} {\bibfnamefont {Ronald}\
  \bibnamefont {Hanson}},\ }\bibfield  {title} {\enquote {\bibinfo {title}
  {Quantum internet: A vision for the road ahead},}\ }\href {\doibase
  10.1126/science.aam9288} {\bibfield  {journal} {\bibinfo  {journal}
  {Science}\ }\textbf {\bibinfo {volume} {362}} (\bibinfo {year} {2018}),\
  10.1126/science.aam9288}\BibitemShut {NoStop}%
\bibitem [{\citenamefont {Benjamin}\ \emph {et~al.}(2009)\citenamefont
  {Benjamin}, \citenamefont {Lovett},\ and\ \citenamefont
  {Smith}}]{Benjamin2009pmb}%
  \BibitemOpen
  \bibfield  {author} {\bibinfo {author} {\bibfnamefont {Simon~C.}\
  \bibnamefont {Benjamin}}, \bibinfo {author} {\bibfnamefont {Brendon~W.}\
  \bibnamefont {Lovett}}, \ and\ \bibinfo {author} {\bibfnamefont {Jason~M.}\
  \bibnamefont {Smith}},\ }\bibfield  {title} {\enquote {\bibinfo {title}
  {Prospects for measurement-based quantum computing with solid state spins},}\
  }\href {\doibase 10.1002/lpor.200810051} {\bibfield  {journal} {\bibinfo
  {journal} {Laser \& Photonics Reviews}\ }\textbf {\bibinfo {volume} {3}},\
  \bibinfo {pages} {556--574} (\bibinfo {year} {2009})}\BibitemShut {NoStop}%
\bibitem [{\citenamefont {Meyer}\ \emph {et~al.}(2004)\citenamefont {Meyer},
  \citenamefont {Alves}, \citenamefont {Hofmann}, \citenamefont {Kriegseis},
  \citenamefont {Forster}, \citenamefont {Bertram}, \citenamefont {Christen},
  \citenamefont {Hoffmann}, \citenamefont {Straßburg}, \citenamefont
  {Dworzak}, \citenamefont {Haboeck},\ and\ \citenamefont
  {Rodina}}]{Meyer2004bed}%
  \BibitemOpen
  \bibfield  {author} {\bibinfo {author} {\bibfnamefont {Bruno~K.}\
  \bibnamefont {Meyer}}, \bibinfo {author} {\bibfnamefont {H.}~\bibnamefont
  {Alves}}, \bibinfo {author} {\bibfnamefont {D.~M.}\ \bibnamefont {Hofmann}},
  \bibinfo {author} {\bibfnamefont {W.}~\bibnamefont {Kriegseis}}, \bibinfo
  {author} {\bibfnamefont {D.}~\bibnamefont {Forster}}, \bibinfo {author}
  {\bibfnamefont {F.}~\bibnamefont {Bertram}}, \bibinfo {author} {\bibfnamefont
  {J.}~\bibnamefont {Christen}}, \bibinfo {author} {\bibfnamefont
  {A.}~\bibnamefont {Hoffmann}}, \bibinfo {author} {\bibfnamefont
  {M.}~\bibnamefont {Straßburg}}, \bibinfo {author} {\bibfnamefont
  {M.}~\bibnamefont {Dworzak}}, \bibinfo {author} {\bibfnamefont
  {U.}~\bibnamefont {Haboeck}}, \ and\ \bibinfo {author} {\bibfnamefont
  {A.~V.}\ \bibnamefont {Rodina}},\ }\bibfield  {title} {\enquote {\bibinfo
  {title} {Bound exciton and donor-acceptor pair recombinations in {ZnO}},}\
  }\href {\doibase 10.1002/pssb.200301962} {\bibfield  {journal} {\bibinfo
  {journal} {Physica Status Solidi (B)}\ }\textbf {\bibinfo {volume} {241}},\
  \bibinfo {pages} {231--260} (\bibinfo {year} {2004})}\BibitemShut {NoStop}%
\bibitem [{\citenamefont {Khaetskii}\ and\ \citenamefont
  {Nazarov}(2001)}]{Khaetskii2001sft}%
  \BibitemOpen
  \bibfield  {author} {\bibinfo {author} {\bibfnamefont {Alexander~V.}\
  \bibnamefont {Khaetskii}}\ and\ \bibinfo {author} {\bibfnamefont {Yuli~V.}\
  \bibnamefont {Nazarov}},\ }\bibfield  {title} {\enquote {\bibinfo {title}
  {Spin-flip transitions between {Zeeman} sublevels in semiconductor quantum
  dots},}\ }\href {\doibase 10.1103/PhysRevB.64.125316} {\bibfield  {journal}
  {\bibinfo  {journal} {Phys. Rev. B}\ }\textbf {\bibinfo {volume} {64}},\
  \bibinfo {pages} {125316} (\bibinfo {year} {2001})}\BibitemShut {NoStop}%
\bibitem [{\citenamefont {Ladd}\ \emph {et~al.}(2010)\citenamefont {Ladd},
  \citenamefont {Jelezko}, \citenamefont {Laflamme}, \citenamefont {Nakamura},
  \citenamefont {Monroe},\ and\ \citenamefont {O'Brien}}]{Ladd2010qc}%
  \BibitemOpen
  \bibfield  {author} {\bibinfo {author} {\bibfnamefont {T.~D.}\ \bibnamefont
  {Ladd}}, \bibinfo {author} {\bibfnamefont {F.}~\bibnamefont {Jelezko}},
  \bibinfo {author} {\bibfnamefont {R.}~\bibnamefont {Laflamme}}, \bibinfo
  {author} {\bibfnamefont {Y.}~\bibnamefont {Nakamura}}, \bibinfo {author}
  {\bibfnamefont {C.}~\bibnamefont {Monroe}}, \ and\ \bibinfo {author}
  {\bibfnamefont {J.~L.}\ \bibnamefont {O'Brien}},\ }\bibfield  {title}
  {\enquote {\bibinfo {title} {Quantum computers},}\ }\href {\doibase
  10.1038/nature08812} {\bibfield  {journal} {\bibinfo  {journal} {Nature}\
  }\textbf {\bibinfo {volume} {464}},\ \bibinfo {pages} {45--53} (\bibinfo
  {year} {2010})}\BibitemShut {NoStop}%
\bibitem [{\citenamefont {Schmidgall}\ \emph {et~al.}(2018)\citenamefont
  {Schmidgall}, \citenamefont {Chakravarthi}, \citenamefont {Gould},
  \citenamefont {Christen}, \citenamefont {Hestroffer}, \citenamefont
  {Hatami},\ and\ \citenamefont {Fu}}]{Schmidgall2018fcs}%
  \BibitemOpen
  \bibfield  {author} {\bibinfo {author} {\bibfnamefont {Emma~R.}\ \bibnamefont
  {Schmidgall}}, \bibinfo {author} {\bibfnamefont {Srivatsa}\ \bibnamefont
  {Chakravarthi}}, \bibinfo {author} {\bibfnamefont {Michael}\ \bibnamefont
  {Gould}}, \bibinfo {author} {\bibfnamefont {Ian~R.}\ \bibnamefont
  {Christen}}, \bibinfo {author} {\bibfnamefont {Karine}\ \bibnamefont
  {Hestroffer}}, \bibinfo {author} {\bibfnamefont {Fariba}\ \bibnamefont
  {Hatami}}, \ and\ \bibinfo {author} {\bibfnamefont {Kai Mei~C.}\ \bibnamefont
  {Fu}},\ }\bibfield  {title} {\enquote {\bibinfo {title} {Frequency control of
  single quantum emitters in integrated photonic circuits},}\ }\href {\doibase
  10.1021/acs.nanolett.7b04717} {\bibfield  {journal} {\bibinfo  {journal}
  {Nano Letters}\ }\textbf {\bibinfo {volume} {18}},\ \bibinfo {pages}
  {1175--1179} (\bibinfo {year} {2018})}\BibitemShut {NoStop}%
\bibitem [{\citenamefont {Schröder}\ \emph {et~al.}(2016)\citenamefont
  {Schröder}, \citenamefont {Mouradian}, \citenamefont {Zheng}, \citenamefont
  {Trusheim}, \citenamefont {Walsh}, \citenamefont {Chen}, \citenamefont {Li},
  \citenamefont {Bayn},\ and\ \citenamefont {Englund}}]{Schroder2016qnd}%
  \BibitemOpen
  \bibfield  {author} {\bibinfo {author} {\bibfnamefont {Tim}\ \bibnamefont
  {Schröder}}, \bibinfo {author} {\bibfnamefont {Sara~L.}\ \bibnamefont
  {Mouradian}}, \bibinfo {author} {\bibfnamefont {Jiabao}\ \bibnamefont
  {Zheng}}, \bibinfo {author} {\bibfnamefont {Matthew~E.}\ \bibnamefont
  {Trusheim}}, \bibinfo {author} {\bibfnamefont {Michael}\ \bibnamefont
  {Walsh}}, \bibinfo {author} {\bibfnamefont {Edward~H.}\ \bibnamefont {Chen}},
  \bibinfo {author} {\bibfnamefont {Luozhou}\ \bibnamefont {Li}}, \bibinfo
  {author} {\bibfnamefont {Igal}\ \bibnamefont {Bayn}}, \ and\ \bibinfo
  {author} {\bibfnamefont {Dirk}\ \bibnamefont {Englund}},\ }\bibfield  {title}
  {\enquote {\bibinfo {title} {Quantum nanophotonics in diamond [invited]},}\
  }\href {\doibase 10.1364/josab.33.000b65} {\bibfield  {journal} {\bibinfo
  {journal} {Journal of the Optical Society of America B}\ }\textbf {\bibinfo
  {volume} {33}},\ \bibinfo {pages} {B65--B83} (\bibinfo {year}
  {2016})}\BibitemShut {NoStop}%
\bibitem [{\citenamefont {Kane}(1998)}]{Kane1998sbn}%
  \BibitemOpen
  \bibfield  {author} {\bibinfo {author} {\bibfnamefont {B~E}\ \bibnamefont
  {Kane}},\ }\bibfield  {title} {\enquote {\bibinfo {title} {A silicon-based
  nuclear spin quantum computer},}\ }\href@noop {} {\bibfield  {journal}
  {\bibinfo  {journal} {Nature}\ }\textbf {\bibinfo {volume} {393}},\ \bibinfo
  {pages} {133--137} (\bibinfo {year} {1998})}\BibitemShut {NoStop}%
\bibitem [{\citenamefont {Ramdas}\ and\ \citenamefont
  {Rodriguez}(1981)}]{Ramdas1981sss}%
  \BibitemOpen
  \bibfield  {author} {\bibinfo {author} {\bibfnamefont {A~K}\ \bibnamefont
  {Ramdas}}\ and\ \bibinfo {author} {\bibfnamefont {S}~\bibnamefont
  {Rodriguez}},\ }\bibfield  {title} {\enquote {\bibinfo {title} {Spectroscopy
  of the solid-state analogues of the hydrogen atom: donors and acceptors in
  semiconductors},}\ }\href {\doibase 10.1088/0034-4885/44/12/002} {\bibfield
  {journal} {\bibinfo  {journal} {Reports on Progress in Physics}\ }\textbf
  {\bibinfo {volume} {44}},\ \bibinfo {pages} {1297--1387} (\bibinfo {year}
  {1981})}\BibitemShut {NoStop}%
\bibitem [{\citenamefont {Linpeng}\ \emph {et~al.}(2016)\citenamefont
  {Linpeng}, \citenamefont {Karin}, \citenamefont {Durnev}, \citenamefont
  {Barbour}, \citenamefont {Glazov}, \citenamefont {Sherman}, \citenamefont
  {Watkins}, \citenamefont {Seto},\ and\ \citenamefont {Fu}}]{Linpeng2016lsr}%
  \BibitemOpen
  \bibfield  {author} {\bibinfo {author} {\bibfnamefont {Xiayu}\ \bibnamefont
  {Linpeng}}, \bibinfo {author} {\bibfnamefont {Todd}\ \bibnamefont {Karin}},
  \bibinfo {author} {\bibfnamefont {M~V}\ \bibnamefont {Durnev}}, \bibinfo
  {author} {\bibfnamefont {Russell}\ \bibnamefont {Barbour}}, \bibinfo {author}
  {\bibfnamefont {M~M}\ \bibnamefont {Glazov}}, \bibinfo {author}
  {\bibfnamefont {E~Ya}\ \bibnamefont {Sherman}}, \bibinfo {author}
  {\bibfnamefont {S~P}\ \bibnamefont {Watkins}}, \bibinfo {author}
  {\bibfnamefont {Satoru}\ \bibnamefont {Seto}}, \ and\ \bibinfo {author}
  {\bibfnamefont {Kai-Mei~C}\ \bibnamefont {Fu}},\ }\bibfield  {title}
  {\enquote {\bibinfo {title} {Longitudinal spin relaxation of donor-bound
  electrons in direct band-gap semiconductors},}\ }\href {\doibase
  10.1103/PhysRevB.94.125401} {\bibfield  {journal} {\bibinfo  {journal}
  {Physical Review B}\ }\textbf {\bibinfo {volume} {94}},\ \bibinfo {pages}
  {125401--125420} (\bibinfo {year} {2016})}\BibitemShut {NoStop}%
\bibitem [{\citenamefont {Sleiter}\ \emph {et~al.}(2013)\citenamefont
  {Sleiter}, \citenamefont {Sanaka}, \citenamefont {Kim}, \citenamefont
  {Lischka}, \citenamefont {Pawlis},\ and\ \citenamefont
  {Yamamoto}}]{Sleiter2013ops}%
  \BibitemOpen
  \bibfield  {author} {\bibinfo {author} {\bibfnamefont {Darin~J.}\
  \bibnamefont {Sleiter}}, \bibinfo {author} {\bibfnamefont {Kaoru}\
  \bibnamefont {Sanaka}}, \bibinfo {author} {\bibfnamefont {Y.~M.}\
  \bibnamefont {Kim}}, \bibinfo {author} {\bibfnamefont {Klaus}\ \bibnamefont
  {Lischka}}, \bibinfo {author} {\bibfnamefont {Alexander}\ \bibnamefont
  {Pawlis}}, \ and\ \bibinfo {author} {\bibfnamefont {Yoshihisa}\ \bibnamefont
  {Yamamoto}},\ }\bibfield  {title} {\enquote {\bibinfo {title} {Optical
  pumping of a single electron spin bound to a fluorine donor in a {ZnSe}
  nanostructure},}\ }\href {\doibase 10.1021/nl303663n} {\bibfield  {journal}
  {\bibinfo  {journal} {Nano Letters}\ }\textbf {\bibinfo {volume} {13}},\
  \bibinfo {pages} {116--120} (\bibinfo {year} {2013})}\BibitemShut {NoStop}%
\bibitem [{\citenamefont {Fu}\ \emph {et~al.}(2005)\citenamefont {Fu},
  \citenamefont {Santori}, \citenamefont {Stanley}, \citenamefont {Holland},\
  and\ \citenamefont {Yamamoto}}]{Fu2005cpt}%
  \BibitemOpen
  \bibfield  {author} {\bibinfo {author} {\bibfnamefont {Kai Mei~C}\
  \bibnamefont {Fu}}, \bibinfo {author} {\bibfnamefont {Charles}\ \bibnamefont
  {Santori}}, \bibinfo {author} {\bibfnamefont {Colin}\ \bibnamefont
  {Stanley}}, \bibinfo {author} {\bibfnamefont {M~C}\ \bibnamefont {Holland}},
  \ and\ \bibinfo {author} {\bibfnamefont {Yoshihisa}\ \bibnamefont
  {Yamamoto}},\ }\bibfield  {title} {\enquote {\bibinfo {title} {Coherent
  population trapping of electron spins in a high-purity n-type {GaAs}
  semiconductor},}\ }\href {\doibase 10.1103/PhysRevLett.95.187405} {\bibfield
  {journal} {\bibinfo  {journal} {Physical Review Letters}\ }\textbf {\bibinfo
  {volume} {95}},\ \bibinfo {pages} {187405--187409} (\bibinfo {year}
  {2005})}\BibitemShut {NoStop}%
\bibitem [{\citenamefont {Santori}\ \emph {et~al.}(2006)\citenamefont
  {Santori}, \citenamefont {Tamarat}, \citenamefont {Neumann}, \citenamefont
  {Wrachtrup}, \citenamefont {Fattal}, \citenamefont {Beausoleil},
  \citenamefont {Rabeau}, \citenamefont {Olivero}, \citenamefont {Greentree},
  \citenamefont {Prawer}, \citenamefont {Jelezko},\ and\ \citenamefont
  {Hemmer}}]{Santori2006cpt}%
  \BibitemOpen
  \bibfield  {author} {\bibinfo {author} {\bibfnamefont {Charles}\ \bibnamefont
  {Santori}}, \bibinfo {author} {\bibfnamefont {Philippe}\ \bibnamefont
  {Tamarat}}, \bibinfo {author} {\bibfnamefont {Philipp}\ \bibnamefont
  {Neumann}}, \bibinfo {author} {\bibfnamefont {Jörg}\ \bibnamefont
  {Wrachtrup}}, \bibinfo {author} {\bibfnamefont {David}\ \bibnamefont
  {Fattal}}, \bibinfo {author} {\bibfnamefont {Raymond~G.}\ \bibnamefont
  {Beausoleil}}, \bibinfo {author} {\bibfnamefont {James}\ \bibnamefont
  {Rabeau}}, \bibinfo {author} {\bibfnamefont {Paolo}\ \bibnamefont {Olivero}},
  \bibinfo {author} {\bibfnamefont {Andrew~D.}\ \bibnamefont {Greentree}},
  \bibinfo {author} {\bibfnamefont {Steven}\ \bibnamefont {Prawer}}, \bibinfo
  {author} {\bibfnamefont {Fedor}\ \bibnamefont {Jelezko}}, \ and\ \bibinfo
  {author} {\bibfnamefont {Philip}\ \bibnamefont {Hemmer}},\ }\bibfield
  {title} {\enquote {\bibinfo {title} {Coherent population trapping of single
  spins in diamond under optical excitation},}\ }\href {\doibase
  10.1103/PhysRevLett.97.247401} {\bibfield  {journal} {\bibinfo  {journal}
  {Physical Review Letters}\ }\textbf {\bibinfo {volume} {97}},\ \bibinfo
  {pages} {247401--247405} (\bibinfo {year} {2006})}\BibitemShut {NoStop}%
\bibitem [{\citenamefont {Gray}\ \emph {et~al.}(1978)\citenamefont {Gray},
  \citenamefont {Whitley},\ and\ \citenamefont {Stroud}}]{Gray1978cta}%
  \BibitemOpen
  \bibfield  {author} {\bibinfo {author} {\bibfnamefont {H~R}\ \bibnamefont
  {Gray}}, \bibinfo {author} {\bibfnamefont {R~M}\ \bibnamefont {Whitley}}, \
  and\ \bibinfo {author} {\bibfnamefont {C~R}\ \bibnamefont {Stroud}},\
  }\bibfield  {title} {\enquote {\bibinfo {title} {Coherent trapping of atomic
  populations},}\ }\href {\doibase 10.1364/OL.3.000218} {\bibfield  {journal}
  {\bibinfo  {journal} {Optics Letters}\ }\textbf {\bibinfo {volume} {3}},\
  \bibinfo {pages} {218--220} (\bibinfo {year} {1978})}\BibitemShut {NoStop}%
\bibitem [{\citenamefont {Xu}\ \emph {et~al.}(2008)\citenamefont {Xu},
  \citenamefont {Sun}, \citenamefont {Berman}, \citenamefont {Steel},
  \citenamefont {Bracker}, \citenamefont {Gammon},\ and\ \citenamefont
  {Sham}}]{Xu2008cpt}%
  \BibitemOpen
  \bibfield  {author} {\bibinfo {author} {\bibfnamefont {Xiaodong}\
  \bibnamefont {Xu}}, \bibinfo {author} {\bibfnamefont {Bo}~\bibnamefont
  {Sun}}, \bibinfo {author} {\bibfnamefont {Paul~R.}\ \bibnamefont {Berman}},
  \bibinfo {author} {\bibfnamefont {Duncan~G.}\ \bibnamefont {Steel}}, \bibinfo
  {author} {\bibfnamefont {Allan~S.}\ \bibnamefont {Bracker}}, \bibinfo
  {author} {\bibfnamefont {Dan}\ \bibnamefont {Gammon}}, \ and\ \bibinfo
  {author} {\bibfnamefont {L.~J.}\ \bibnamefont {Sham}},\ }\bibfield  {title}
  {\enquote {\bibinfo {title} {Coherent population trapping of an electron spin
  in a single negatively charged quantum dot},}\ }\href {\doibase
  10.1038/nphys1054} {\bibfield  {journal} {\bibinfo  {journal} {Nature
  Physics}\ }\textbf {\bibinfo {volume} {4}},\ \bibinfo {pages} {692--695}
  (\bibinfo {year} {2008})}\BibitemShut {NoStop}%
\bibitem [{\citenamefont {Gonzalez}\ \emph {et~al.}(1982)\citenamefont
  {Gonzalez}, \citenamefont {Block}, \citenamefont {Cox},\ and\ \citenamefont
  {Hervé}}]{Gonzalez1982mrs}%
  \BibitemOpen
  \bibfield  {author} {\bibinfo {author} {\bibfnamefont {C.}~\bibnamefont
  {Gonzalez}}, \bibinfo {author} {\bibfnamefont {D.}~\bibnamefont {Block}},
  \bibinfo {author} {\bibfnamefont {R.~T.}\ \bibnamefont {Cox}}, \ and\
  \bibinfo {author} {\bibfnamefont {A.}~\bibnamefont {Hervé}},\ }\bibfield
  {title} {\enquote {\bibinfo {title} {Magnetic resonance studies of shallow
  donors in zinc oxide},}\ }\href {\doibase 10.1016/0022-0248(82)90351-7}
  {\bibfield  {journal} {\bibinfo  {journal} {Journal of Crystal Growth}\
  }\textbf {\bibinfo {volume} {59}},\ \bibinfo {pages} {357--362} (\bibinfo
  {year} {1982})}\BibitemShut {NoStop}%
\bibitem [{\citenamefont {Block}\ \emph {et~al.}(1982)\citenamefont {Block},
  \citenamefont {Herv\'e},\ and\ \citenamefont {Cox}}]{Block1982odm}%
  \BibitemOpen
  \bibfield  {author} {\bibinfo {author} {\bibfnamefont {D}~\bibnamefont
  {Block}}, \bibinfo {author} {\bibfnamefont {A}~\bibnamefont {Herv\'e}}, \
  and\ \bibinfo {author} {\bibfnamefont {R~T}\ \bibnamefont {Cox}},\ }\bibfield
   {title} {\enquote {\bibinfo {title} {Optically detected magnetic resonance
  and optically detected endor of shallow indium donors in {ZnO}},}\ }\href
  {\doibase 10.1103/PhysRevB.25.6049} {\bibfield  {journal} {\bibinfo
  {journal} {Physical Review B}\ }\textbf {\bibinfo {volume} {25}},\ \bibinfo
  {pages} {6049--6052} (\bibinfo {year} {1982})}\BibitemShut {NoStop}%
\bibitem [{\citenamefont {Kumar}\ \emph {et~al.}(2013)\citenamefont {Kumar},
  \citenamefont {Mohammadbeigi}, \citenamefont {Alagha}, \citenamefont {Deng},
  \citenamefont {Anderson}, \citenamefont {Wintschel},\ and\ \citenamefont
  {Watkins}}]{kumar2013oed}%
  \BibitemOpen
  \bibfield  {author} {\bibinfo {author} {\bibfnamefont {E~Senthil}\
  \bibnamefont {Kumar}}, \bibinfo {author} {\bibfnamefont {F}~\bibnamefont
  {Mohammadbeigi}}, \bibinfo {author} {\bibfnamefont {S}~\bibnamefont
  {Alagha}}, \bibinfo {author} {\bibfnamefont {Z~W}\ \bibnamefont {Deng}},
  \bibinfo {author} {\bibfnamefont {I~P}\ \bibnamefont {Anderson}}, \bibinfo
  {author} {\bibfnamefont {T}~\bibnamefont {Wintschel}}, \ and\ \bibinfo
  {author} {\bibfnamefont {S~P}\ \bibnamefont {Watkins}},\ }\bibfield  {title}
  {\enquote {\bibinfo {title} {Optical evidence for donor behavior of {Sb} in
  {ZnO} nanowires},}\ }\href {\doibase 10.1063/1.4799385} {\bibfield  {journal}
  {\bibinfo  {journal} {Applied Physics Letters}\ }\textbf {\bibinfo {volume}
  {102}},\ \bibinfo {pages} {132105} (\bibinfo {year} {2013})}\BibitemShut
  {NoStop}%
\bibitem [{SI()}]{SI}%
  \BibitemOpen
  \href@noop {} {}\bibinfo {note} {See Supplementary Material}\BibitemShut
  {NoStop}%
\bibitem [{\citenamefont {Ding}\ \emph {et~al.}(2009)\citenamefont {Ding},
  \citenamefont {Li}, \citenamefont {He}, \citenamefont {Ge}, \citenamefont
  {Wang}, \citenamefont {Ning}, \citenamefont {Dai}, \citenamefont {Ling},\
  and\ \citenamefont {Xu}}]{Ding2009cbe}%
  \BibitemOpen
  \bibfield  {author} {\bibinfo {author} {\bibfnamefont {L.}~\bibnamefont
  {Ding}}, \bibinfo {author} {\bibfnamefont {B.~K.}\ \bibnamefont {Li}},
  \bibinfo {author} {\bibfnamefont {H.~T.}\ \bibnamefont {He}}, \bibinfo
  {author} {\bibfnamefont {W.~K.}\ \bibnamefont {Ge}}, \bibinfo {author}
  {\bibfnamefont {J.~N.}\ \bibnamefont {Wang}}, \bibinfo {author}
  {\bibfnamefont {J.~Q.}\ \bibnamefont {Ning}}, \bibinfo {author}
  {\bibfnamefont {X.~M.}\ \bibnamefont {Dai}}, \bibinfo {author} {\bibfnamefont
  {C.~C.}\ \bibnamefont {Ling}}, \ and\ \bibinfo {author} {\bibfnamefont
  {S.~J.}\ \bibnamefont {Xu}},\ }\bibfield  {title} {\enquote {\bibinfo {title}
  {Classification of bound exciton complexes in bulk {ZnO} by
  magnetophotoluminescence spectroscopy},}\ }\href {\doibase 10.1063/1.3087762}
  {\bibfield  {journal} {\bibinfo  {journal} {Journal of Applied Physics}\
  }\textbf {\bibinfo {volume} {105}},\ \bibinfo {pages} {053511} (\bibinfo
  {year} {2009})}\BibitemShut {NoStop}%
\bibitem [{\citenamefont {Wagner}\ \emph {et~al.}(2009)\citenamefont {Wagner},
  \citenamefont {Schulze}, \citenamefont {Kirste}, \citenamefont {Cobet},
  \citenamefont {Hoffmann}, \citenamefont {Rauch}, \citenamefont {Rodina},
  \citenamefont {Meyer}, \citenamefont {Röder},\ and\ \citenamefont
  {Thonke}}]{wagner2009gvb}%
  \BibitemOpen
  \bibfield  {author} {\bibinfo {author} {\bibfnamefont {Markus~R.}\
  \bibnamefont {Wagner}}, \bibinfo {author} {\bibfnamefont {Jan~Hindrik}\
  \bibnamefont {Schulze}}, \bibinfo {author} {\bibfnamefont {Ronny}\
  \bibnamefont {Kirste}}, \bibinfo {author} {\bibfnamefont {Munise}\
  \bibnamefont {Cobet}}, \bibinfo {author} {\bibfnamefont {Axel}\ \bibnamefont
  {Hoffmann}}, \bibinfo {author} {\bibfnamefont {Christian}\ \bibnamefont
  {Rauch}}, \bibinfo {author} {\bibfnamefont {Anna~V.}\ \bibnamefont {Rodina}},
  \bibinfo {author} {\bibfnamefont {Bruno~K.}\ \bibnamefont {Meyer}}, \bibinfo
  {author} {\bibfnamefont {Uwe}\ \bibnamefont {Röder}}, \ and\ \bibinfo
  {author} {\bibfnamefont {Klaus}\ \bibnamefont {Thonke}},\ }\bibfield  {title}
  {\enquote {\bibinfo {title} {${\ensuremath{\Gamma}}_{7}$ valence band
  symmetry related hole fine splitting of bound excitons in {ZnO} observed in
  magneto-optical studies},}\ }\href {\doibase 10.1103/PhysRevB.80.205203}
  {\bibfield  {journal} {\bibinfo  {journal} {Physical Review B}\ }\textbf
  {\bibinfo {volume} {80}},\ \bibinfo {pages} {205203--205209} (\bibinfo {year}
  {2009})}\BibitemShut {NoStop}%
\bibitem [{\citenamefont {Wischmeier}\ \emph {et~al.}(2006)\citenamefont
  {Wischmeier}, \citenamefont {Voss}, \citenamefont {Rückmann}, \citenamefont
  {Gutowski}, \citenamefont {Mofor}, \citenamefont {Bakin},\ and\ \citenamefont
  {Waag}}]{Wischmeier2006dse}%
  \BibitemOpen
  \bibfield  {author} {\bibinfo {author} {\bibfnamefont {L.}~\bibnamefont
  {Wischmeier}}, \bibinfo {author} {\bibfnamefont {T.}~\bibnamefont {Voss}},
  \bibinfo {author} {\bibfnamefont {I.}~\bibnamefont {Rückmann}}, \bibinfo
  {author} {\bibfnamefont {J.}~\bibnamefont {Gutowski}}, \bibinfo {author}
  {\bibfnamefont {A.~C.}\ \bibnamefont {Mofor}}, \bibinfo {author}
  {\bibfnamefont {A.}~\bibnamefont {Bakin}}, \ and\ \bibinfo {author}
  {\bibfnamefont {A.}~\bibnamefont {Waag}},\ }\bibfield  {title} {\enquote
  {\bibinfo {title} {Dynamics of surface-excitonic emission in {ZnO}
  nanowires},}\ }\href {\doibase 10.1103/PhysRevB.74.195333} {\bibfield
  {journal} {\bibinfo  {journal} {Physical Review B}\ }\textbf {\bibinfo
  {volume} {74}},\ \bibinfo {pages} {195333--195342} (\bibinfo {year}
  {2006})}\BibitemShut {NoStop}%
\bibitem [{\citenamefont {Biswas}\ \emph {et~al.}(2011)\citenamefont {Biswas},
  \citenamefont {Jung}, \citenamefont {Kim}, \citenamefont {Kumar},
  \citenamefont {Hughes}, \citenamefont {Newcomb}, \citenamefont {Henry},\ and\
  \citenamefont {McGlynn}}]{Biswas2011mos}%
  \BibitemOpen
  \bibfield  {author} {\bibinfo {author} {\bibfnamefont {Mahua}\ \bibnamefont
  {Biswas}}, \bibinfo {author} {\bibfnamefont {Yun~Suk}\ \bibnamefont {Jung}},
  \bibinfo {author} {\bibfnamefont {Hong~Koo}\ \bibnamefont {Kim}}, \bibinfo
  {author} {\bibfnamefont {Kumarappan}\ \bibnamefont {Kumar}}, \bibinfo
  {author} {\bibfnamefont {Gregory~J.}\ \bibnamefont {Hughes}}, \bibinfo
  {author} {\bibfnamefont {S.}~\bibnamefont {Newcomb}}, \bibinfo {author}
  {\bibfnamefont {Martin~O.}\ \bibnamefont {Henry}}, \ and\ \bibinfo {author}
  {\bibfnamefont {Enda}\ \bibnamefont {McGlynn}},\ }\bibfield  {title}
  {\enquote {\bibinfo {title} {Microscopic origins of the surface exciton
  photoluminescence peak in {ZnO} nanostructures},}\ }\href {\doibase
  10.1103/PhysRevB.83.235320} {\bibfield  {journal} {\bibinfo  {journal}
  {Physical Review B}\ }\textbf {\bibinfo {volume} {83}},\ \bibinfo {pages}
  {235320--235330} (\bibinfo {year} {2011})}\BibitemShut {NoStop}%
\bibitem [{\citenamefont {Grabowska}\ \emph {et~al.}(2005)\citenamefont
  {Grabowska}, \citenamefont {Meaney}, \citenamefont {Nanda}, \citenamefont
  {Mosnier}, \citenamefont {Henry}, \citenamefont {Duclère},\ and\
  \citenamefont {McGlynn}}]{Grabowska2005see}%
  \BibitemOpen
  \bibfield  {author} {\bibinfo {author} {\bibfnamefont {J.}~\bibnamefont
  {Grabowska}}, \bibinfo {author} {\bibfnamefont {A.}~\bibnamefont {Meaney}},
  \bibinfo {author} {\bibfnamefont {K.~K.}\ \bibnamefont {Nanda}}, \bibinfo
  {author} {\bibfnamefont {J.~P.}\ \bibnamefont {Mosnier}}, \bibinfo {author}
  {\bibfnamefont {M.~O.}\ \bibnamefont {Henry}}, \bibinfo {author}
  {\bibfnamefont {J.~R.}\ \bibnamefont {Duclère}}, \ and\ \bibinfo {author}
  {\bibfnamefont {E.}~\bibnamefont {McGlynn}},\ }\bibfield  {title} {\enquote
  {\bibinfo {title} {Surface excitonic emission and quenching effects in {ZnO}
  nanowire/nanowall systems: Limiting effects on device potential},}\ }\href
  {\doibase 10.1103/PhysRevB.71.115439} {\bibfield  {journal} {\bibinfo
  {journal} {Physical Review B}\ }\textbf {\bibinfo {volume} {71}},\ \bibinfo
  {pages} {115439--115446} (\bibinfo {year} {2005})}\BibitemShut {NoStop}%
\bibitem [{\citenamefont {Travnikov}\ \emph {et~al.}(1989)\citenamefont
  {Travnikov}, \citenamefont {Freiberg},\ and\ \citenamefont
  {Savikhin}}]{Travnikov1989sez}%
  \BibitemOpen
  \bibfield  {author} {\bibinfo {author} {\bibfnamefont {V.V.}\ \bibnamefont
  {Travnikov}}, \bibinfo {author} {\bibfnamefont {A.}~\bibnamefont {Freiberg}},
  \ and\ \bibinfo {author} {\bibfnamefont {S.F.}\ \bibnamefont {Savikhin}},\
  }\bibfield  {title} {\enquote {\bibinfo {title} {Surface excitons in {ZnO}
  crystals},}\ }\href {\doibase 10.1088/0953-8984/1/5/002} {\bibfield
  {journal} {\bibinfo  {journal} {Journal of Physics: Condensed Matter}\
  }\textbf {\bibinfo {volume} {1}},\ \bibinfo {pages} {847--854} (\bibinfo
  {year} {1989})}\BibitemShut {NoStop}%
\bibitem [{\citenamefont {Reimer}\ \emph {et~al.}(2016)\citenamefont {Reimer},
  \citenamefont {Bulgarini}, \citenamefont {Fognini}, \citenamefont {Heeres},
  \citenamefont {Witek}, \citenamefont {Versteegh}, \citenamefont {Rubino},
  \citenamefont {Braun}, \citenamefont {Kamp}, \citenamefont {Höfling},
  \citenamefont {Dalacu}, \citenamefont {Lapointe}, \citenamefont {Poole},\
  and\ \citenamefont {Zwiller}}]{Reimer2016opb}%
  \BibitemOpen
  \bibfield  {author} {\bibinfo {author} {\bibfnamefont {M.~E.}\ \bibnamefont
  {Reimer}}, \bibinfo {author} {\bibfnamefont {G.}~\bibnamefont {Bulgarini}},
  \bibinfo {author} {\bibfnamefont {A.}~\bibnamefont {Fognini}}, \bibinfo
  {author} {\bibfnamefont {R.~W.}\ \bibnamefont {Heeres}}, \bibinfo {author}
  {\bibfnamefont {B.~J.}\ \bibnamefont {Witek}}, \bibinfo {author}
  {\bibfnamefont {M.~A.M.}\ \bibnamefont {Versteegh}}, \bibinfo {author}
  {\bibfnamefont {A.}~\bibnamefont {Rubino}}, \bibinfo {author} {\bibfnamefont
  {T.}~\bibnamefont {Braun}}, \bibinfo {author} {\bibfnamefont
  {M.}~\bibnamefont {Kamp}}, \bibinfo {author} {\bibfnamefont {S.}~\bibnamefont
  {Höfling}}, \bibinfo {author} {\bibfnamefont {D.}~\bibnamefont {Dalacu}},
  \bibinfo {author} {\bibfnamefont {J.}~\bibnamefont {Lapointe}}, \bibinfo
  {author} {\bibfnamefont {P.~J.}\ \bibnamefont {Poole}}, \ and\ \bibinfo
  {author} {\bibfnamefont {V.}~\bibnamefont {Zwiller}},\ }\bibfield  {title}
  {\enquote {\bibinfo {title} {Overcoming power broadening of the quantum dot
  emission in a pure wurtzite nanowire},}\ }\href {\doibase
  10.1103/PhysRevB.93.195316} {\bibfield  {journal} {\bibinfo  {journal}
  {Physical Review B}\ }\textbf {\bibinfo {volume} {93}},\ \bibinfo {pages}
  {195316} (\bibinfo {year} {2016})}\BibitemShut {NoStop}%
\bibitem [{\citenamefont {Holmes}\ \emph {et~al.}(2015)\citenamefont {Holmes},
  \citenamefont {Kako}, \citenamefont {Choi}, \citenamefont {Arita},\ and\
  \citenamefont {Arakawa}}]{Holmes2015sdi}%
  \BibitemOpen
  \bibfield  {author} {\bibinfo {author} {\bibfnamefont {M.}~\bibnamefont
  {Holmes}}, \bibinfo {author} {\bibfnamefont {S.}~\bibnamefont {Kako}},
  \bibinfo {author} {\bibfnamefont {K.}~\bibnamefont {Choi}}, \bibinfo {author}
  {\bibfnamefont {M.}~\bibnamefont {Arita}}, \ and\ \bibinfo {author}
  {\bibfnamefont {Y.}~\bibnamefont {Arakawa}},\ }\bibfield  {title} {\enquote
  {\bibinfo {title} {Spectral diffusion and its influence on the emission
  linewidths of site-controlled {GaN} nanowire quantum dots},}\ }\href
  {\doibase 10.1103/PhysRevB.92.115447} {\bibfield  {journal} {\bibinfo
  {journal} {Physical Review B}\ }\textbf {\bibinfo {volume} {92}},\ \bibinfo
  {pages} {115447--115454} (\bibinfo {year} {2015})}\BibitemShut {NoStop}%
\bibitem [{\citenamefont {Empedocles}\ and\ \citenamefont
  {Bawendi}(1999)}]{Empedocles1999isd}%
  \BibitemOpen
  \bibfield  {author} {\bibinfo {author} {\bibfnamefont {S.~A.}\ \bibnamefont
  {Empedocles}}\ and\ \bibinfo {author} {\bibfnamefont {M.~G.}\ \bibnamefont
  {Bawendi}},\ }\bibfield  {title} {\enquote {\bibinfo {title} {Influence of
  spectral diffusion on the line shapes of single {CdSe} nanocrystallite
  quantum dots},}\ }\href {\doibase 10.1021/jp983305x} {\bibfield  {journal}
  {\bibinfo  {journal} {Journal of Physical Chemistry B}\ }\textbf {\bibinfo
  {volume} {103}},\ \bibinfo {pages} {1826--1830} (\bibinfo {year}
  {1999})}\BibitemShut {NoStop}%
\bibitem [{\citenamefont {MacQuarrie}\ \emph {et~al.}(2021)\citenamefont
  {MacQuarrie}, \citenamefont {Chartrand}, \citenamefont {Higginbottom},
  \citenamefont {Morse}, \citenamefont {Karasyuk}, \citenamefont {Roorda},\
  and\ \citenamefont {Simmons}}]{MacQuarrie2021gtc}%
  \BibitemOpen
  \bibfield  {author} {\bibinfo {author} {\bibfnamefont {Evan}\ \bibnamefont
  {MacQuarrie}}, \bibinfo {author} {\bibfnamefont {Camille}\ \bibnamefont
  {Chartrand}}, \bibinfo {author} {\bibfnamefont {Daniel}\ \bibnamefont
  {Higginbottom}}, \bibinfo {author} {\bibfnamefont {Kevin}\ \bibnamefont
  {Morse}}, \bibinfo {author} {\bibfnamefont {Valentin}\ \bibnamefont
  {Karasyuk}}, \bibinfo {author} {\bibfnamefont {Sjoerd}\ \bibnamefont
  {Roorda}}, \ and\ \bibinfo {author} {\bibfnamefont {Stephanie}\ \bibnamefont
  {Simmons}},\ }\bibfield  {title} {\enquote {\bibinfo {title} {Generating {T}
  centres in photonic silicon-on-insulator material by ion implantation},}\
  }\href {\doibase 10.1088/1367-2630/ac291f} {\bibfield  {journal} {\bibinfo
  {journal} {New Journal of Physics}\ }\textbf {\bibinfo {volume} {23}},\
  \bibinfo {pages} {103008} (\bibinfo {year} {2021})}\BibitemShut {NoStop}%
\bibitem [{\citenamefont {Linpeng}\ \emph {et~al.}(2021)\citenamefont
  {Linpeng}, \citenamefont {Karin}, \citenamefont {Durnev}, \citenamefont
  {Glazov}, \citenamefont {Schott}, \citenamefont {Wieck}, \citenamefont
  {Ludwig},\ and\ \citenamefont {Fu}}]{Linpeng2021osc}%
  \BibitemOpen
  \bibfield  {author} {\bibinfo {author} {\bibfnamefont {Xiayu}\ \bibnamefont
  {Linpeng}}, \bibinfo {author} {\bibfnamefont {Todd}\ \bibnamefont {Karin}},
  \bibinfo {author} {\bibfnamefont {Mikhail~V.}\ \bibnamefont {Durnev}},
  \bibinfo {author} {\bibfnamefont {Mikhail~M.}\ \bibnamefont {Glazov}},
  \bibinfo {author} {\bibfnamefont {Rüdiger}\ \bibnamefont {Schott}}, \bibinfo
  {author} {\bibfnamefont {Andreas~D.}\ \bibnamefont {Wieck}}, \bibinfo
  {author} {\bibfnamefont {Arne}\ \bibnamefont {Ludwig}}, \ and\ \bibinfo
  {author} {\bibfnamefont {Kai Mei~C.}\ \bibnamefont {Fu}},\ }\bibfield
  {title} {\enquote {\bibinfo {title} {Optical spin control and coherence
  properties of acceptor bound holes in strained {GaAs}},}\ }\href {\doibase
  10.1103/PhysRevB.103.115412} {\bibfield  {journal} {\bibinfo  {journal}
  {Physical Review B}\ }\textbf {\bibinfo {volume} {103}},\ \bibinfo {pages}
  {115412--115422} (\bibinfo {year} {2021})}\BibitemShut {NoStop}%
\bibitem [{\citenamefont {Kumar}\ \emph {et~al.}(2016)\citenamefont {Kumar},
  \citenamefont {Mohammadbeigi}, \citenamefont {Boatner},\ and\ \citenamefont
  {Watkins}}]{Kumar2016hrp}%
  \BibitemOpen
  \bibfield  {author} {\bibinfo {author} {\bibfnamefont {E.~Senthil}\
  \bibnamefont {Kumar}}, \bibinfo {author} {\bibfnamefont {F.}~\bibnamefont
  {Mohammadbeigi}}, \bibinfo {author} {\bibfnamefont {L.~A.}\ \bibnamefont
  {Boatner}}, \ and\ \bibinfo {author} {\bibfnamefont {S.~P.}\ \bibnamefont
  {Watkins}},\ }\bibfield  {title} {\enquote {\bibinfo {title} {High-resolution
  photoluminescence spectroscopy of {Sn}-doped {ZnO} single crystals},}\ }\href
  {\doibase 10.1016/j.jlumin.2016.01.028} {\bibfield  {journal} {\bibinfo
  {journal} {Journal of Luminescence}\ }\textbf {\bibinfo {volume} {176}},\
  \bibinfo {pages} {47--51} (\bibinfo {year} {2016})}\BibitemShut {NoStop}%
\end{thebibliography}%


\begin{thebibliography}{3}%
\makeatletter
\providecommand \@ifxundefined [1]{%
 \@ifx{#1\undefined}
}%
\providecommand \@ifnum [1]{%
 \ifnum #1\expandafter \@firstoftwo
 \else \expandafter \@secondoftwo
 \fi
}%
\providecommand \@ifx [1]{%
 \ifx #1\expandafter \@firstoftwo
 \else \expandafter \@secondoftwo
 \fi
}%
\providecommand \natexlab [1]{#1}%
\providecommand \enquote  [1]{``#1''}%
\providecommand \bibnamefont  [1]{#1}%
\providecommand \bibfnamefont [1]{#1}%
\providecommand \citenamefont [1]{#1}%
\providecommand \href@noop [0]{\@secondoftwo}%
\providecommand \href [0]{\begingroup \@sanitize@url \@href}%
\providecommand \@href[1]{\@@startlink{#1}\@@href}%
\providecommand \@@href[1]{\endgroup#1\@@endlink}%
\providecommand \@sanitize@url [0]{\catcode `\\12\catcode `\$12\catcode
  `\&12\catcode `\#12\catcode `\^12\catcode `\_12\catcode `\%12\relax}%
\providecommand \@@startlink[1]{}%
\providecommand \@@endlink[0]{}%
\providecommand \url  [0]{\begingroup\@sanitize@url \@url }%
\providecommand \@url [1]{\endgroup\@href {#1}{\urlprefix }}%
\providecommand \urlprefix  [0]{URL }%
\providecommand \Eprint [0]{\href }%
\providecommand \doibase [0]{https://doi.org/}%
\providecommand \selectlanguage [0]{\@gobble}%
\providecommand \bibinfo  [0]{\@secondoftwo}%
\providecommand \bibfield  [0]{\@secondoftwo}%
\providecommand \translation [1]{[#1]}%
\providecommand \BibitemOpen [0]{}%
\providecommand \bibitemStop [0]{}%
\providecommand \bibitemNoStop [0]{.\EOS\space}%
\providecommand \EOS [0]{\spacefactor3000\relax}%
\providecommand \BibitemShut  [1]{\csname bibitem#1\endcsname}%
\let\auto@bib@innerbib\@empty
\bibitem [{\citenamefont {Meyer}\ \emph {et~al.}(2004)\citenamefont {Meyer},
  \citenamefont {Alves}, \citenamefont {Hofmann}, \citenamefont {Kriegseis},
  \citenamefont {Forster}, \citenamefont {Bertram}, \citenamefont {Christen},
  \citenamefont {Hoffmann}, \citenamefont {Straßburg}, \citenamefont
  {Dworzak}, \citenamefont {Haboeck},\ and\ \citenamefont
  {Rodina}}]{Meyer2004bed}%
  \BibitemOpen
  \bibfield  {author} {\bibinfo {author} {\bibfnamefont {B.~K.}\ \bibnamefont
  {Meyer}}, \bibinfo {author} {\bibfnamefont {H.}~\bibnamefont {Alves}},
  \bibinfo {author} {\bibfnamefont {D.~M.}\ \bibnamefont {Hofmann}}, \bibinfo
  {author} {\bibfnamefont {W.}~\bibnamefont {Kriegseis}}, \bibinfo {author}
  {\bibfnamefont {D.}~\bibnamefont {Forster}}, \bibinfo {author} {\bibfnamefont
  {F.}~\bibnamefont {Bertram}}, \bibinfo {author} {\bibfnamefont
  {J.}~\bibnamefont {Christen}}, \bibinfo {author} {\bibfnamefont
  {A.}~\bibnamefont {Hoffmann}}, \bibinfo {author} {\bibfnamefont
  {M.}~\bibnamefont {Straßburg}}, \bibinfo {author} {\bibfnamefont
  {M.}~\bibnamefont {Dworzak}}, \bibinfo {author} {\bibfnamefont
  {U.}~\bibnamefont {Haboeck}},\ and\ \bibinfo {author} {\bibfnamefont {A.~V.}\
  \bibnamefont {Rodina}},\ }\href {https://doi.org/10.1002/pssb.200301962}
  {\bibfield  {journal} {\bibinfo  {journal} {Physica Status Solidi (B)}\
  }\textbf {\bibinfo {volume} {241}},\ \bibinfo {pages} {231} (\bibinfo {year}
  {2004})}\BibitemShut {NoStop}%
\bibitem [{\citenamefont {Wagner}\ \emph {et~al.}(2009)\citenamefont {Wagner},
  \citenamefont {Schulze}, \citenamefont {Kirste}, \citenamefont {Cobet},
  \citenamefont {Hoffmann}, \citenamefont {Rauch}, \citenamefont {Rodina},
  \citenamefont {Meyer}, \citenamefont {Röder},\ and\ \citenamefont
  {Thonke}}]{wagner2009gvb}%
  \BibitemOpen
  \bibfield  {author} {\bibinfo {author} {\bibfnamefont {M.~R.}\ \bibnamefont
  {Wagner}}, \bibinfo {author} {\bibfnamefont {J.~H.}\ \bibnamefont {Schulze}},
  \bibinfo {author} {\bibfnamefont {R.}~\bibnamefont {Kirste}}, \bibinfo
  {author} {\bibfnamefont {M.}~\bibnamefont {Cobet}}, \bibinfo {author}
  {\bibfnamefont {A.}~\bibnamefont {Hoffmann}}, \bibinfo {author}
  {\bibfnamefont {C.}~\bibnamefont {Rauch}}, \bibinfo {author} {\bibfnamefont
  {A.~V.}\ \bibnamefont {Rodina}}, \bibinfo {author} {\bibfnamefont {B.~K.}\
  \bibnamefont {Meyer}}, \bibinfo {author} {\bibfnamefont {U.}~\bibnamefont
  {Röder}},\ and\ \bibinfo {author} {\bibfnamefont {K.}~\bibnamefont
  {Thonke}},\ }\href {https://doi.org/10.1103/PhysRevB.80.205203} {\bibfield
  {journal} {\bibinfo  {journal} {Physical Review B}\ }\textbf {\bibinfo
  {volume} {80}},\ \bibinfo {pages} {205203} (\bibinfo {year}
  {2009})}\BibitemShut {NoStop}%
\bibitem [{\citenamefont {Ding}\ \emph {et~al.}(2009)\citenamefont {Ding},
  \citenamefont {Li}, \citenamefont {He}, \citenamefont {Ge}, \citenamefont
  {Wang}, \citenamefont {Ning}, \citenamefont {Dai}, \citenamefont {Ling},\
  and\ \citenamefont {Xu}}]{Ding2009cbe}%
  \BibitemOpen
  \bibfield  {author} {\bibinfo {author} {\bibfnamefont {L.}~\bibnamefont
  {Ding}}, \bibinfo {author} {\bibfnamefont {B.~K.}\ \bibnamefont {Li}},
  \bibinfo {author} {\bibfnamefont {H.~T.}\ \bibnamefont {He}}, \bibinfo
  {author} {\bibfnamefont {W.~K.}\ \bibnamefont {Ge}}, \bibinfo {author}
  {\bibfnamefont {J.~N.}\ \bibnamefont {Wang}}, \bibinfo {author}
  {\bibfnamefont {J.~Q.}\ \bibnamefont {Ning}}, \bibinfo {author}
  {\bibfnamefont {X.~M.}\ \bibnamefont {Dai}}, \bibinfo {author} {\bibfnamefont
  {C.~C.}\ \bibnamefont {Ling}},\ and\ \bibinfo {author} {\bibfnamefont
  {S.~J.}\ \bibnamefont {Xu}},\ }\href {https://doi.org/10.1063/1.3087762}
  {\bibfield  {journal} {\bibinfo  {journal} {Journal of Applied Physics}\
  }\textbf {\bibinfo {volume} {105}},\ \bibinfo {pages} {053511} (\bibinfo
  {year} {2009})}\BibitemShut {NoStop}%
\end{thebibliography}%

\end{document}